\documentclass[draft]{AGUJournal}
\draftfalse

\newcommand{\mvec}[1]{\mathbf{#1}}

\journalname{Geophysical Research Letters}
\usepackage{graphics,graphicx}
\usepackage{cancel}

\begin{document}

%
%


\title{Global Ten-Moment Multifluid Simulations of the Solar Wind Interaction with Mercury: From the Planetary Conducting Core to the Dynamic Magnetosphere}

%
%




\authors{Chuanfei Dong,\affil{1,2}  Liang Wang,\affil{1,2}  Ammar Hakim,\affil{2}  Amitava Bhattacharjee,\affil{1,2}  James A. Slavin,\affil{3}  Gina A. DiBraccio,\affil{4} Kai Germaschewski\affil{5}}

\affiliation{1}{Department of Astrophysical Sciences, Princeton University, Princeton, New Jersey, USA}

\affiliation{2}{Princeton Center for Heliophysics, Princeton Plasma Physics Laboratory, Princeton University, Princeton, New Jersey, USA}

\affiliation{3}{Department of Climate and Space Sciences and Engineering, University of Michigan, Ann Arbor, Michigan, USA}

\affiliation{4}{NASA Goddard Space Flight Center, Greenbelt, Maryland, USA}

\affiliation{5}{Space Science Center and Physics Department, University of New Hampshire, Durham, New Hampshire, USA}






\correspondingauthor{Chuanfei Dong}{dcfy@princeton.edu}




\begin{keypoints}
\item The new model can reproduce observations beyond MHD including dawn-dusk asymmetries in Mercury's magnetotail and field-aligned currents
\item The new model is essential for capturing the electron physics associated with collisionless magnetic reconnection in Mercury's magnetosphere
\item The induction response arising from the electromagnetically-coupled interior plays an important role in solar wind-Mercury interaction
\end{keypoints}

%
%


\begin{abstract}
For the first time, we explore the tightly coupled interior-magnetosphere system of Mercury by employing a three-dimensional ten-moment multifluid model. This novel fluid model incorporates the non-ideal effects including the Hall effect, inertia, and tensorial pressures that are critical for collisionless magnetic reconnection; therefore, it is particularly well suited for investigating \emph{collisionless} magnetic reconnection in Mercury's magnetotail and at the planet's magnetopause. The model is able to reproduce the observed magnetic field vectors, field-aligned currents, and cross-tail current sheet asymmetry (beyond the MHD approach) and the simulation results are in good agreement with spacecraft observations. We also study the magnetospheric response of Mercury to a hypothetical extreme event with an enhanced solar wind dynamic pressure, which demonstrates the significance of induction effects resulting from the electromagnetically-coupled interior. More interestingly, plasmoids (or flux ropes) are formed in Mercury's magnetotail during the event, indicating the highly dynamic nature of Mercury's magnetosphere. 
\end{abstract}

\section{Introduction}\label{sec:Intro}

Mercury, the closest planet to the Sun, is the only terrestrial planet other than Earth that possesses an intrinsic global magnetic field \citep{ness74,ness75}. The recent MErcury Surface, Space ENvironment, GEochemistry, and Ranging (MESSENGER) mission to Mercury presented us with the first opportunity to explore this planet's magnetosphere in great detail since the brief flybys of Mariner 10 \citep[e.g.,][]{solomon07,slavin07}. Many Earth-like magnetospheric features were observed at Mercury, including, but not limited to, magnetopause reconnection \citep{slavin09,dibraccio13}, the concomitant flux transfer events (FTEs) \citep{slavin12} and cusp plasma filaments \citep{slavin14,pho16}, magnetotail flux ropes or plasmoids \citep{dibraccio15}, substorm processes including tail loading-unloading \citep{imber17}, plasma wave activities \citep{sun15}, dipolarization fronts \citep{sundberg12} and the associated electron acceleration \citep{dewey17}, cross-tail current sheet asymmetry and substorm current wedge formation \citep{poh17}, field-aligned currents \citep{anderson14}, and Kelvin-Helmholtz vortices \citep{sundberg10,liljeblad14,gershman15}.

According to MESSENGER observations, Mercury's dipole moment is much weaker than that of Earth, only 195 nT R$_M^3$ (where R$_M$ is Mercury's radius, 2440 km), and is offset in the northward direction by 484 $\pm$ 11 km or $\approx$ 0.2 R$_M$ \citep{anderson11}. Later, those values were slightly modified in \citet{anderson12}. Due to the relatively weak intrinsic planetary magnetic moment and the most extreme solar wind driving forces in the solar system, Mercury has a small but extremely dynamic magnetosphere whose size is about 5\% that of Earth's magnetosphere \citep{winslow13}. More interestingly, Mercury has a large electrically conductive iron core with a radius of $\approx$ 0.8 R$_M$ \citep{smith12,hauck13}. A unique aspect of Mercury's interaction system is that the large conducting core can induce observable magnetic fields in Mercury's magnetosphere \citep{slavin14,zhong15,johnson16}. It is worth noting that \citet{hood79} and \citet{grosser04} made some early quantitive estimates of the induction effect at Mercury. The core-induced magnetic fields have been demonstrated to play an important role in Mercury's global solar wind interaction, especially during extreme space weather events \citep{slavin14,jia15,heyner16,slavin19}. While the induction response generates additional magnetic flux that may protect Mercury from solar wind erosion, magnetic reconnection between the interplanetary magnetic field (IMF) and the planetary field removes magnetic flux from the dayside magnetopause and enables transfer of energy and momentum to the planetary inner magnetosphere, which consequently leads to the direct entry of solar wind plasma into the system. The magnetic flux transferred to the nightside magnetosphere may immediately undergo reconnection or be stored and later returned to the dayside during an intense episode of reconnection in the tail \citep{slavin14}. Magnetotail reconnection is also the dominant plasma process that transfers energy and momentum into Mercury's inner tail region by converting stored magnetic energy in the tail lobe into plasma kinetic energy in the plasma sheet. Magnetic reconnection, therefore, plays a crucial role in manipulating the magnetospheric dynamics of Mercury and other planets in our solar system and beyond.

Despite the significant achievements accomplished by direct spacecraft observations, \emph{in situ} measurements are often taken at limited points along the trajectories of orbits or flybys. Such limitations, however, can be alleviated by numerical simulations, which allow the interpretation of \emph{in situ} measurements in a three-dimensional context and distinguishing temporal from spatial fluctuations as well. Thus, numerical models, combined with \emph{in situ} data, are the key for providing a global description of solar wind-planet interaction. In recent years, our understanding of terrestrial bodies has been significantly advanced by increasingly sophisticated numerical models. A large number of global models based on either fluid or hybrid (kinetic ion particles and massless electron fluid) approach have been developed for both magnetized planets such as Mercury \citep[e.g.,][]{kabin08,kidder08,pavel10,muller12,richer12,jia15,exner18} and unmagnetized planets such as Mars \citep{ma14,dong14,dong15b,dong18b,dong18c,modolo16,ledvina17} as well as exoplanets \citep{johansson11,dong17a,dong17b,dong18a,dong19}. However, none of these global models can accurately treat collisionless magnetic reconnection due to their lack of detailed electron physics. In order to solve this issue with affordable computational costs, two broad approaches have been proposed. \citet{toth16} studied Ganymede's magnetosphere by employing a Hall magnetohydrodynamic model with embedded particle-in-cell boxes (MHD-EPIC) such that they can capture the collisionless reconnection physics in prescribed local regions. Meanwhile, \citet{wang18} developed a novel ten-moment multifluid model to study Ganymede's magnetosphere. Other than relying on the prescribed local PIC boxes, the new global multi-moment multifluid model incorporating the higher-order moments is capable of reproducing some critical aspects of the reconnection physics from PIC simulations \citep{wang15,ng15,ng17,ng18}. 

Until now, no such approach (i.e., either MHD-EPIC or the multi-moment multifluid approach) has been applied to study Mercury. This work will, therefore, be the first study of Mercury's dynamic magnetosphere using a ten-moment multifluid model. In order to capture the induction effects arising from the interior-magnetosphere electromagnetic coupling, we also implemented a resistive mantle and an electrically conductive core inside Mercury in this new model. This paper is structured as follows. In Section \ref{sec:Model}, the ten-moment multifluid model and the model setup for Mercury are described. In Section \ref{sec:data-model}, we first validate the model through data-model comparison with MESSENGER data and then discuss the model results. We also conduct a hypothetical extreme event case study to demonstrate the significance of the induction effects. The conclusion is given in Section \ref{sec:conclusion}.

\section{Ten-Moment Multifluid Model for Mercury} \label{sec:Model}

\subsection{Ten-Moment Equations} \label{subsec:ModelEqs}
In this section, we briefly introduce the ten-moment multifluid model for Mercury within the \textsc{Gkeyll} framework\footnote{gkeyll.rtfd.io}. The ten moments refer to mass density $mn$, momentum $mn u_x$, $mn u_y$, $mn u_z$ and pressure tensor ${P}_{xx}$, ${P}_{xy}$, ${P}_{xz}$, ${P}_{yy}$, ${P}_{yz}$, ${P}_{zz}$. Conceptually, the ten-moment model is akin to a fluid version of particle-in-cell (PIC) code, truncated at a certain order of moment, i.e., second-order moment, the pressure. For Mercury, we solve ten-moment equations for both protons and electrons. It is noteworthy that the ten-moment model has been employed to study magnetic reconnection in multi-species plasmas including O$^+$, H$^+$, and e$^-$ \citep{dong16}. The ten-moment equations for each species are given as follows:
 
\begin{eqnarray}
\frac{\partial\left(m_{s}n_{s}\right)}{\partial t}+\frac{\partial\left(m_{s}n_{s}u_{i,s}\right)}{\partial x_{i}} & = & 0,\\
\frac{\partial\left(m_{s}n_{s}u_{i,s}\right)}{\partial t}+\frac{\partial\mathcal{P}_{ij,s}}{\partial x_{j}} & = & n_{s}q_{s}\left(E_{i}+\epsilon_{ijk}u_{j,s}B_{k}\right),\label{eq:10m-momentum}\\
\frac{\partial\mathcal{P}_{ij,s}}{\partial t}+\frac{\partial\mathcal{Q}_{ijk,s}}{\partial x_{k}} & = & n_{s}q_{s}u_{[i,s}E_{j]}+\frac{q_{s}}{m_{s}}\epsilon_{[ikl}\mathcal{P}_{kj,s]}B_{l}.\label{eq:10m-pressure}\\ \nonumber
\end{eqnarray}
where $q$ is the charge, $E$ and $B$ are electric field and magnetic field, respectively. The subscripts $s=e,i$ represent the electrons and ion species. It will be neglected hereinafter for convenience. The square brackets in Equation (\ref{eq:10m-pressure}) surrounding the indices represent the minimal sum over permutations of free indices needed to yield completely symmetric tensors. The first-order moment is defined as $m n u_{i}\equiv m \int f v_{i}d\mathbf{v}$, where $f$ is the phase space distribution function, $m$ and $v_i$ denote the individual particle mass and velocity, respectively. Similarly, the second-order moment, $\mathcal{P}_{ij}$, and third-order moment, $\mathcal{Q}_{ijk}$, are defined as
\begin{eqnarray}
 & \mathcal{P}_{ij} & =m\int fv_{i}v_{j}d\mathbf{v}\nonumber \\
 &  & =m\int f\left(v_{i}-u_{i}\right)\left(v_{j}-u_{j}\right)d\mathbf{v}+nmu_{i}u_{j}\nonumber \\
 &  & =P_{ij}+nmu_{i}u_{j}.
\end{eqnarray}
and,
\begin{eqnarray}
 & \mathcal{Q}_{ijk} & =m\int fv_{i}v_{j}v_{k}d\mathbf{v}\nonumber \\
 &  & =m\int f\left(v_{i}-u_{i}\right)\left(v_{j}-u_{j}\right)\left(v_{k}-u_{k}\right)d\mathbf{v}+u_{[i}\mathcal{P}_{jk]}-2nmu_{i}u_{j}u_{k}\nonumber \\
 &  & =Q_{ijk}+u_{[i}\mathcal{P}_{jk]}-2nmu_{i}u_{j}u_{k}
\end{eqnarray}
where $P_{ij}$ is the pressure tensor and ${Q}_{ijk}$ is the heat flux tensor. One of the key issues for a multi-moment multifluid model is the closure problem, i.e., how to close the equation systems and incorporate kinetic effects into a fluid framework, which is still an active research topic in fluid dynamics and plasma physics \citep{hunana18}. In this work, we adopt the following 3D closure simplified by \citet{wang15} based on Landau-fluid closures \citep[e.g.,][]{hammett90}:
\begin{eqnarray}
\partial_{m}Q_{ijm}\approx v_{t}\left|k\right|\left(P_{ij}-p\delta_{ij}\right).
\end{eqnarray}
where $v_t$ refers to the local thermal speed, $p$ is the scalar pressure, and $k$ is a free parameter that effectively allows for deviations from isotropy at length scales less than $1/|k|$. For \emph{collisionless} magnetic reconnection, $k$ should be a function of $d_e$ given that \emph{collisionless} magnetic reconnection takes place on the length scale of electron inertial lengths, $d_{e}$. Following the work of \citet{wang18}, we define $k_{s}(\mathbf{x},t)$ as $10/d_{s}(\mathbf{x},t)$, where $d_{s}(\mathbf{x},t)$ is the local inertial length of species $s$ as a function of $\mathbf{x}$ and $t$, such that it can provide a more accurate heat flux approximation because the species inertial length for the Mercury system can vary greatly in space. Interestingly, such closure can well reproduce the \emph{collisionless} reconnection physics from a fully kinetic particle-in-cell code as shown in \citet{wang15}.

The electromagnetic field is solved by full Maxwell equations
\begin{eqnarray}
\frac{1}{c^{2}}\frac{\partial\mathbf{E}}{\partial t} & = & \nabla\times\mathbf{B}-\mu_{0}\mathbf{J},\label{eq:maxwell-dE} \\
\frac{\partial\mathbf{B}}{\partial t} & = & -\nabla\times\mathbf{E},
\end{eqnarray}
where $\mathbf{J}$ is the electric current density. Inside the planet interior $J = \sigma \mathbf{E}$, where plasma convection, $\mathbf{u}$, can be neglected. Unlike the traditional magnetohydrodynamic (MHD) or hybrid models that solve the electric field $\mathbf{E}$ by Ohm's law, here we update $\mathbf{E}$ directly through the Ampere's law, Equation (\ref{eq:maxwell-dE}). Therefore, electromagnetic waves are fully supported, similar to a PIC code. In order to demonstrate how a ten-moment model supports the reconnection electric field in \emph{collisionless} magnetospheres, we rearrange Equation (\ref{eq:10m-momentum}) and obtain the following generalized Ohm's law \citep[e.g.][]{wang15,lin2017}:
\begin{eqnarray}
\mathbf{E}+\mathbf{v}\times\mathbf{B} &=& \underbrace{\cancel{\eta\mathbf{J}}}_{0} + \frac{\mathbf{J}\times\mathbf{B}}{n |e|}-\frac{\nabla\cdot\mathbf{P}_{e}}{n |e|} 
+\frac{m_{e}}{n |e|^{2}}\left[\frac{\partial\mathbf{J}}{\partial t}+\nabla\cdot\left(\mathbf{v}\mathbf{J}+\mathbf{J}\mathbf{v}-\frac{\mathbf{J}\mathbf{J}}{n |e|}\right)\right].\label{eq:ohm}
\end{eqnarray}
It should be noted that the Ohm's law formulated above is not numerically solved in the model.

In the case of 2D anti-parallel magnetic reconnection without a guild field, $\mathbf{B}=0$ (hence $\mathbf{v} \times \mathbf{B}=0$ and $\mathbf{J} \times \mathbf{B}=0$) at reconnection sites or X-points, therefore the divergence of the electron pressure tensor and the total derivative of the electric current are the primary sources of the reconnection electric field in a \emph{collisionless} ($\eta=0$) system (see Equation \ref{eq:ohm} or \citet{zweibel09}). It is further demonstrated by PIC simulations that the reconnection electric field, $E_z$, is largely supported by the divergence of the off-diagonal
elements of $\mathbf{P_{e}}$, i.e., $E_z = -(\partial_x P_{xz,e} + \partial_y P_{yz,e})/n_e |e|$, while traditional MHD and hybrid models only assume a scalar pressure, which does not contribute to $E_z$ at reconnection sites \citep{wang15}. Even if a guide field exists, one can still  get the similar conclusion. The multi-moment multifluid code has been used to study many laboratory and space plasma physics problems \citep[e.g.,][]{ng15,ng19,wang18,tenBarge19}. The details concerning the numerics and benchmark examples have been described in \citet{hakim06}, \citet{hakim08} and \citet{wang19}.

\subsection{Model Setup for Mercury}\label{subsec:ModelSetup}

In a ten-moment model, the time step is mainly restricted by the speed of light. For this reason, we relax this restriction by using an artificially reduced speed of light, $c = 3000$ km/s. We also apply a reduced ion-electron mass ratio $m_i/m_e$ = 25 as the previous study \citep{wang18}, which is sufficiently large to separate the electron and ion scales. The upstream ion inertial length is set to $d_{i,in} = 0.05 R_M$ and electron inertial length $d_{e,in} = 0.01 R_M$. We adopt the Mercury-Solar-Orbital (MSO) coordinates, where the $x$ axis points from Mercury toward the Sun, the $z$ axis is perpendicular to planet's orbital plane, and the $y$ axis completes the right-hand system. The computational domain is defined by $ - 15 R_M \le x \le 5 R_M$, $- 30 R_M \le y,z \le 30 R_M$ with a nonuniform stretched Cartesian grid. The smallest grid size is 0.01 $R_M$, and in turn, five cells are employed to resolve the ion inertial length and one cell for the electron inertial length. In order to capture the magnetospheric physics with minimum influences from numerical resistivity, we use a total of $\sim 4\times 10^9$ cells such that we are able to cover most of the Hermean magnetosphere with the finest grid mesh (i.e., 0.01 $R_M$ resolution). 

We implement Mercury's intrinsic dipole magnetic field $\mathbf{B_0}$ with an equatorial surface strength of 195 nT and centered at (0, 0, 0.2 R$_M$) in MSO. The dipole field is prescribed and fixed in time. The total magnetic field $\mathbf{B}$ equals $\mathbf{B_0} + \mathbf{B_1}$, and we only solve the perturbation magnetic field, $\mathbf{B_1}$, in the model. The inner boundary for electromagnetic fields is set at core surface (0.8 R$_M$) where the conducting wall boundary conditions are applied. For plasma fluids, the inner boundary is set at the planet's surface, such that fluid moment equations are not solved inside the planet. If the surface plasma flow has an inflow component (i.e., $\mathbf{u} \cdot \mathbf{r} < 0$), absorbing boundary conditions are applied. If the surface plasma flow has an outflow component (i.e., $\mathbf{u} \cdot \mathbf{r} > 0$), we set the radial velocity equal to zero, and the plasma density and pressure are fixed at 1 cm$^{-3}$ and 0.001 nPa, respectively \citep{jia15}. Outer boundary conditions are inflow at $x = 5 R_M$ and float at the flanks and tail side.

\section{Results and Discussion}\label{sec:data-model}

In this section, we first validate the model through data-model comparison. We then discuss the model results including day- and night-side magnetic reconnection, field-aligned currents, and cross-tail current sheet asymmetry. Finally, we present Mercury's magnetospheric response to a hypothetical extreme event.

\subsection{Model Validation through Data-Model Comparison}

When magnetic reconnection occurs at the dayside magnetopause, it leads to an efficient transfer of energy and flux from the solar wind into the magnetosphere, which ultimately drives reconnection in the magnetotail. We choose to study MESSENGER's second flyby on October 6, 2018 (hereinafter referred to as M2), during which the IMF had a southward (negative $B_z$) component. For M2, the solar wind parameters are as follows: solar wind density,  40 $cm^{-3}$, solar wind velocity in MSO, $\left(-400,50,0\right)$ km/s, solar wind temperature 18 eV, and IMF in MSO, $\left(-15.2,8.4,-8.5\right)$ nT, where the y-component of the solar wind flow velocity results from Mercury's orbital motion \citep{jia15}.

\begin{figure*}[!ht]
\centering\
\includegraphics[width=0.66\textwidth]{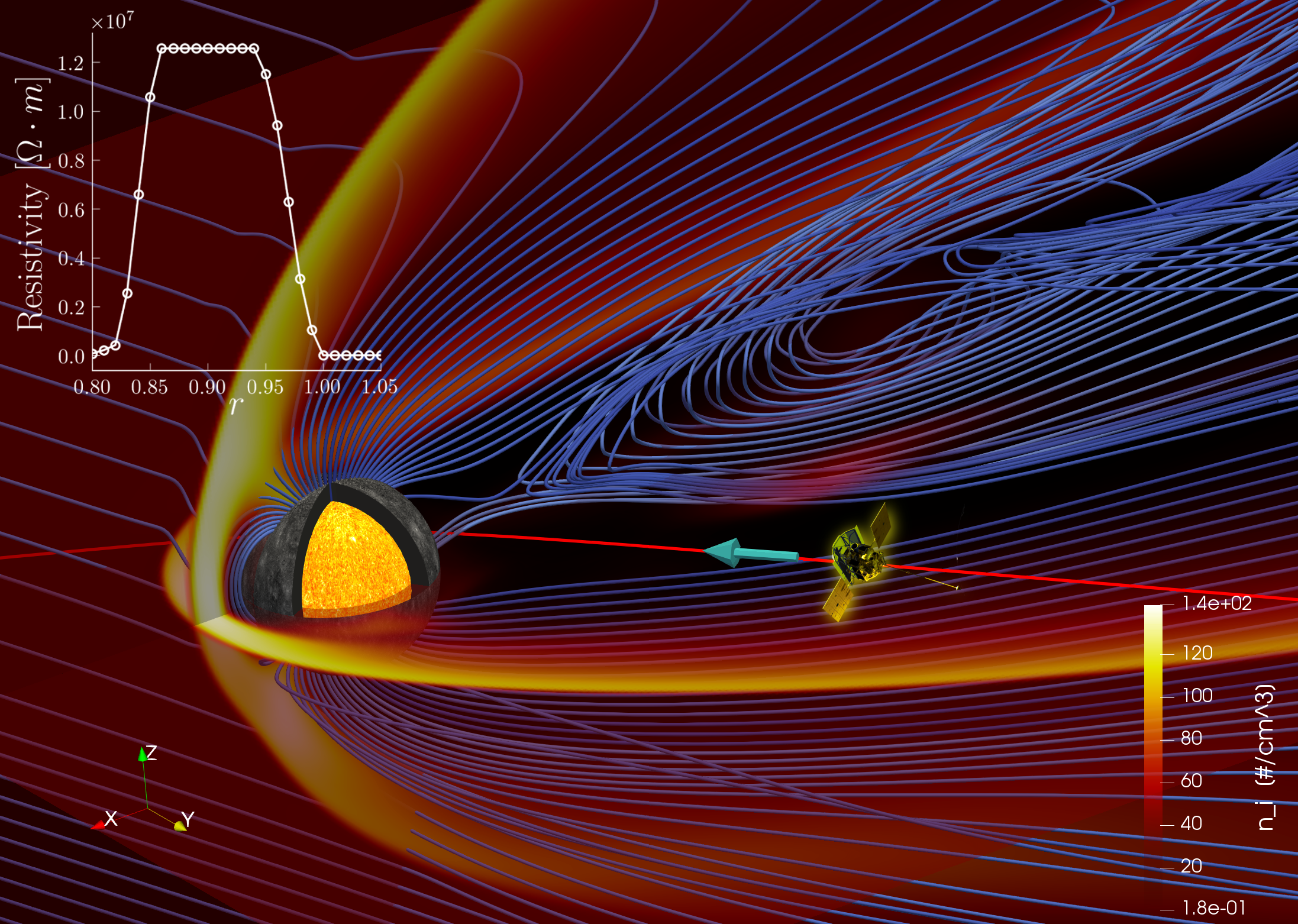}
\includegraphics[width=0.68\textwidth]{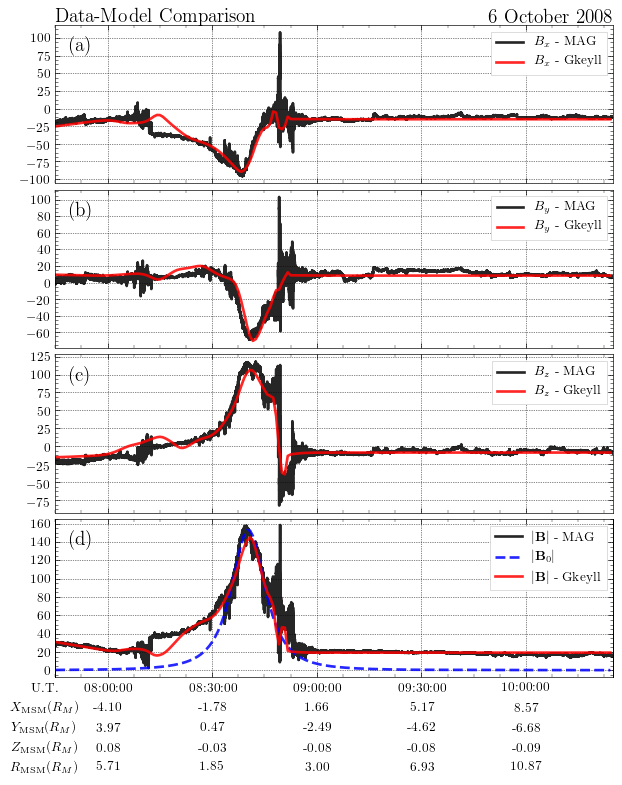}
\caption{Top: Mercury's three-dimensional magnetosphere from the ten-moment multifluid calculation. The color contours depict the ion density in cm$^{-3}$. The ``hot'' sphere inside Mercury represents its conducting core with a size R$_{c}$ = 0.8 R$_{M}$. The magnetic field lines are presented in blue. The red curve together with a cyan arrow represents MESSENGER's M2 trajectory. The radial resistivity profile adopted from \citet{jia15} is shown at the top-left corner. Bottom: Data-model comparison of magnetic fields along MESSENGER's M2 trajectory.}
\label{3DMS}
\end{figure*}

Figure \ref{3DMS} (top) presents Mercury's three-dimensional magnetosphere from the ten-moment multifluid calculation. Magnetospheric characteristics such as the bow shock, magnetosheath, magnetopause, and magnetotail are clearly captured. In detail, the ``hot'' sphere (0.8 R$_M$) inside Mercury represents Mercury's electrically conductive core. M2 trajectory is plotted in red, pointing from night/dusk side to day/dawn side and near Mercury's equatorial plane. Between the conducting core and planet's surface, there exists a highly resistive mantle. The radial resistivity profile shown in the top-left corner of Figure \ref{3DMS} has been adopted from \citet{jia15}, and the white dots in the embedding plot are the grid points used in the model, i.e., 0.01 $R_M$.

To validate our model calculations, we compare the simulation results with MESSENGER's magnetic field data. Panels (a)-(d) of Figure \ref{3DMS} compare the model-calculated magnetic field components along M2 (in red) to MESSENGER magnetometer measurements (in black). Mercury's (unperturbed) intrinsic dipole magnetic field is also plotted as a reference (the blue dashed line in the last row) to illustrate how the global solar wind interaction affects Mercury's magnetosphere. Good agreement is observed between the model calculations and MESSENGER observations in Figure \ref{3DMS}, thus ensuring the validity of our novel approach. Due to the lack of accurate solar wind measurements, we are not able to reproduce the FTE (i.e., the spike structure at 08:50 UTC) observed by MESSENGER. As will be shown below, our model is capable of reproducing other important MESSENGER observations (beyond the MHD approach); therefore our numerical study by adopting this new model represents a crucial step toward establishing a modeling framework that enables self-consistent characterization of Mercury's tightly coupled interior-magnetosphere system.

\subsection{Model Results Analysis and Discussion}

\subsubsection{Dawn-Dusk Asymmetries in Mercury's Magnetotail and Field-Aligned Currents}

\begin{figure}[!htbp]
\centering\
\includegraphics[width=1.0\textwidth,angle=0]{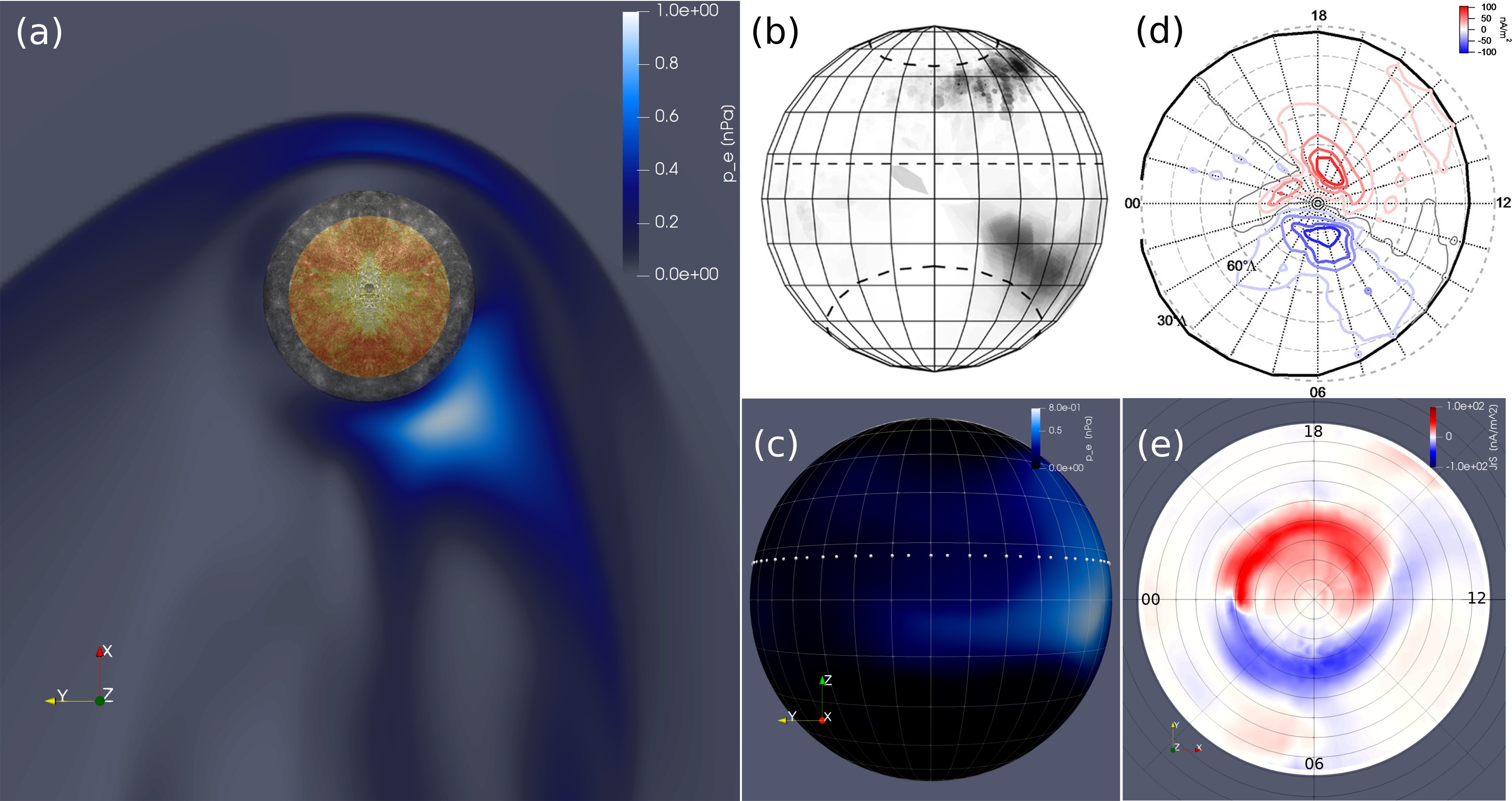}
\caption{(a) Electron pressure ($p_e$) distribution in Mercury's magnetic equatorial plane at z = 0.2 R$_{M}$. (b) X-Ray Spectrometer (XRS) observations of energetic electron-induced surface fluorescence at Mercury's nightside surface from \citet{lindsay16}. (c) Electron pressure ($p_e$) distribution at Mercury's nightside surface from the ten-moment model. (d) Contour plot of radial current density, $J_{rS}$, at Mercury's (northern hemisphere) surface displayed versus local time in hours from \citet{anderson14} based on MESSENGER magnetometer observations. (e) Calculated radial current density, $J_{rS}$, at Mercury's (northern hemisphere) surface from the ten-moment model.}
\label{2DAsym}
\end{figure}

Dawn-dusk asymmetry is a ubiquitous phenomenon in planetary magnetotails. Notably, the ten-moment multifluid model is able to capture the remarkable asymmetry exhibited in Mercury's magnetotail current sheet. Figure \ref{2DAsym}(a) depicts the electron pressure scalar ($p_e$) in Mercury's magnetic equatorial plane (at z = 0.2 R$_{M}$ in MSO), where the cross-tail current sheet is located. From Figure \ref{2DAsym}(a), one can see that (1) more hot electrons are present at the dawnside especially in the inner tail region, and (2) the asymmetry in $p_e$ gradually decreases with increasing distance down the tail. By analyzing the simulation results, we find a slightly dawnward preference in magnetotail reconnection, however, the dawn-dusk asymmetry of the x-line is not significant, probably due to the lack of a dominant amount of Na$^+$ on the duskside as suggested by \citet{poh17}. Here, we conclude that the exhibited asymmetry in hot electron distribution is caused by the dual effect of Mercury's magnetotail reconnection and the dawnward drifts of electrons. When approaching Mercury, the kinetic energy of the sunward reconnection outflow can be easily converted to thermal energy due to the tailward pressure gradient force, leading to more notable asymmetry near the planet relative to the far tail. Meanwhile, the sunward electron flow also drifts to dawnside according to the perpendicular drift velocity of species $s$, $\mathbf{u}_{s\perp}$, derived from the cross product of Equation (\ref{eq:10m-momentum}) and $\mathbf{B}$,
\begin{equation}\label{drifts} 
  \mvec{u}_{s\perp} =
  \frac{\mvec{E}\times\mvec{B}}{B^2}
  -\frac{\nabla\cdot\mvec{P}_s \times\mvec{B}}{q_s n_s B^2}
  -\frac{m_s}{q_s B^2} \frac{d\mvec{u}_s}{dt}\times\mvec{B}
\end{equation}
where the first term is the $\mathbf{E}\times\mathbf{B}$ drift, the second term incorporates the diamagnetic drift and the curvature drift (given $\mvec{P}_s = \mvec{I} p_{s\perp} + \mvec{b}\mvec{b} (p_{s\parallel}-p_{s\perp})  + \mvec{\Pi}_s$, where $\mvec{\Pi}_s$ is the off-diagonal part of the pressure tensor), while the last term contains the polarization drift. Interestingly, an asymmetry also manifests in the X-ray fluorescence (XRF) from MESSENGER X-Ray Spectrometer (XRS) observations at Mercury's nightside surface (Figure \ref{2DAsym}(b)). It is noteworthy that the calculated electron pressure, $p_e$, at Mercury's nightside surface (Figure \ref{2DAsym}(c)) depicts similar patterns as the XRF, supporting the idea of electron-induced surface fluorescence by \citet{lindsay16}.

In addition to the asymmetries, we also present the simulation results for the field-aligned currents (or Birkeland currents) at Mercury's northern hemisphere surface in Figure \ref{2DAsym}(e). The model predicts that the currents flow downward (in blue) at dawn and upward (in red) at dusk, which are consistent with MESSENGER observations shown in Figure \ref{2DAsym}(d) and analogous to Region 1 (R1) Birkeland currents at Earth. More importantly, our simulation results for the current density values at the planetary surface also agree well with MESSENGER observations. MESSENGER magnetic field data show that the maximum and minimum $J_{rS}$ are $\pm115 nA/m^2$ \citep{anderson14}, and in comparison, the calculated maximum and minimum values from our model are 115 $nA/m^2$ and -150 $nA/m^2$, respectively.

\subsubsection{Magnetotail and Magnetopause Reconnection}

\begin{figure}[!htbp]
\centering\
\includegraphics[width=1.0\textwidth]{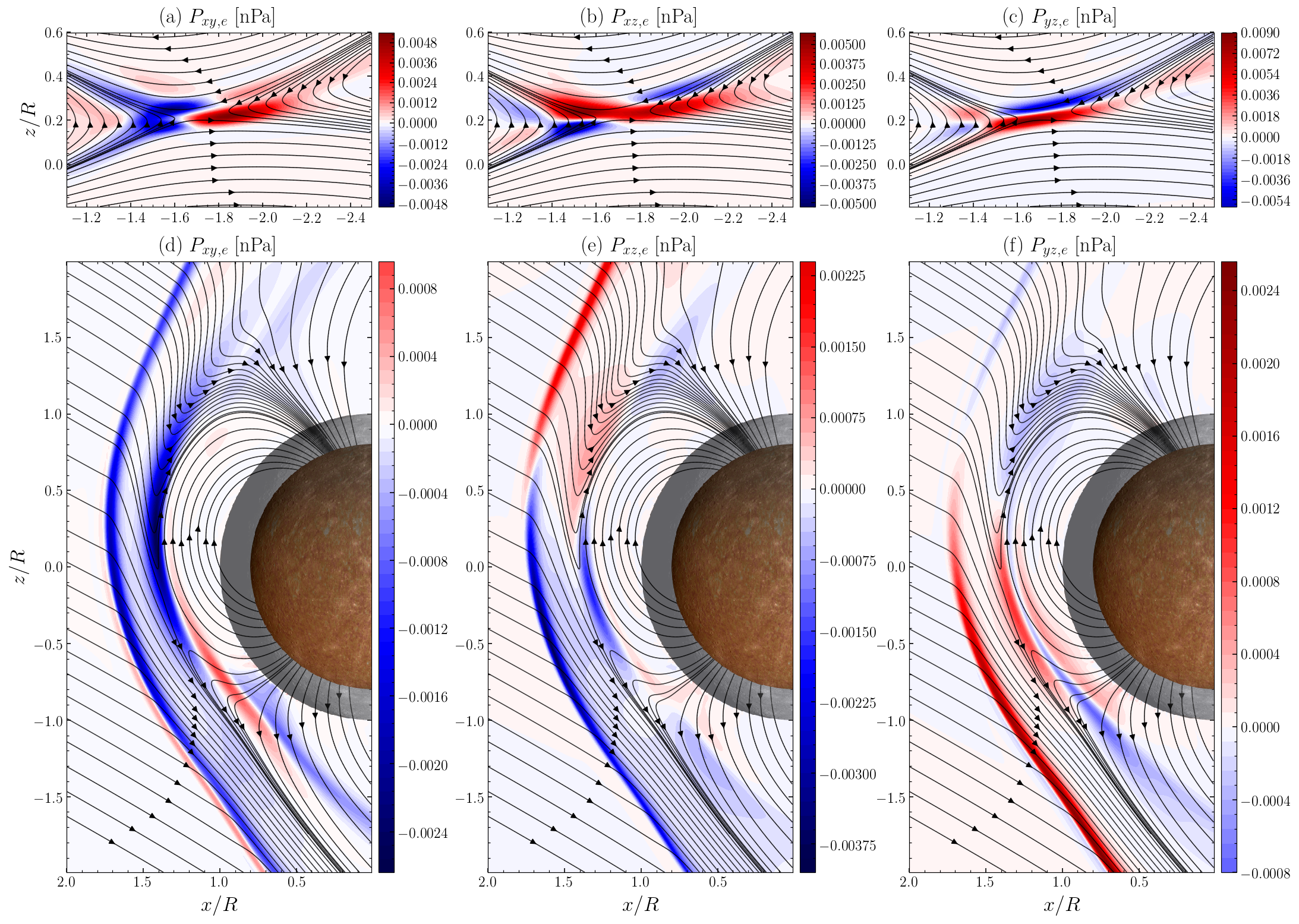}
\caption{Magnetic reconnection in Mercury's magnetotail (first row) and at the magnetopause (second row). Different components of the electron pressure tensor off-diagonal terms ($P_{xy,e}$, $P_{xz,e}$ and $P_{yz,e}$ in nP) are plotted.}
\label{NightDayRec}
\end{figure}

In order to demonstrate that the magnetic reconnection in our calculations is driven by detailed electron physics instead of numerical dissipation as in \citet{jia15,jia19}, we further study the magnetic reconnection in Mercury's magnetotail and at the planet's magnetopause. We first investigate the magnetotail reconnection where the electron reconnection physics is less contaminated given that the tail is less affected by direct solar wind interaction than Mercury's dayside magnetopause. Note that previous full PIC simulations showed that the divergence of the off-diagonal elements of electron pressure tensor, $\mathbf{P_e}$, is the main source of the reconnection electric field \citep{wang15,wilson16}, which can be verified from Equation (\ref{eq:ohm}) as well. We therefore plot $P_{xy,e}$, $P_{xz,e}$ and $P_{yz,e}$ in the first row of Figure \ref{NightDayRec}. Among the three $\mathbf{P_e}$ off-diagonal terms, $P_{yz,e}$ has the largest amplitude and gradient, therefore is the most important term, consistent with previous studies \citep[e.g.,][]{wang15,divin16,wang18}.

Subsequently, we investigated the magnetopause reconnection. Again the three $\mathbf{P_e}$ off-diagonal elements are shown in the second row of Figure \ref{NightDayRec}, where the reconnection rate ranges from 0.08 to 0.2, depending on the locations. In comparison with Figure \ref{NightDayRec}(a-c), Figure \ref{NightDayRec}(d-f) also exhibits different patterns for the $\mathbf{P_e}$ off-diagonal elements. In addition to the reconnection physics, Figure \ref{NightDayRec} clearly depicts the magnetopause location ($\approx$1.4 R$_M$) and the bow shock location ($\approx$1.8 R$_M$), consistent with the previous validated study by \citet{jia15}.

\subsubsection{Extreme Event Case Study}

\begin{figure}[!htbp]
\centering\
\includegraphics[width=0.9\textwidth]{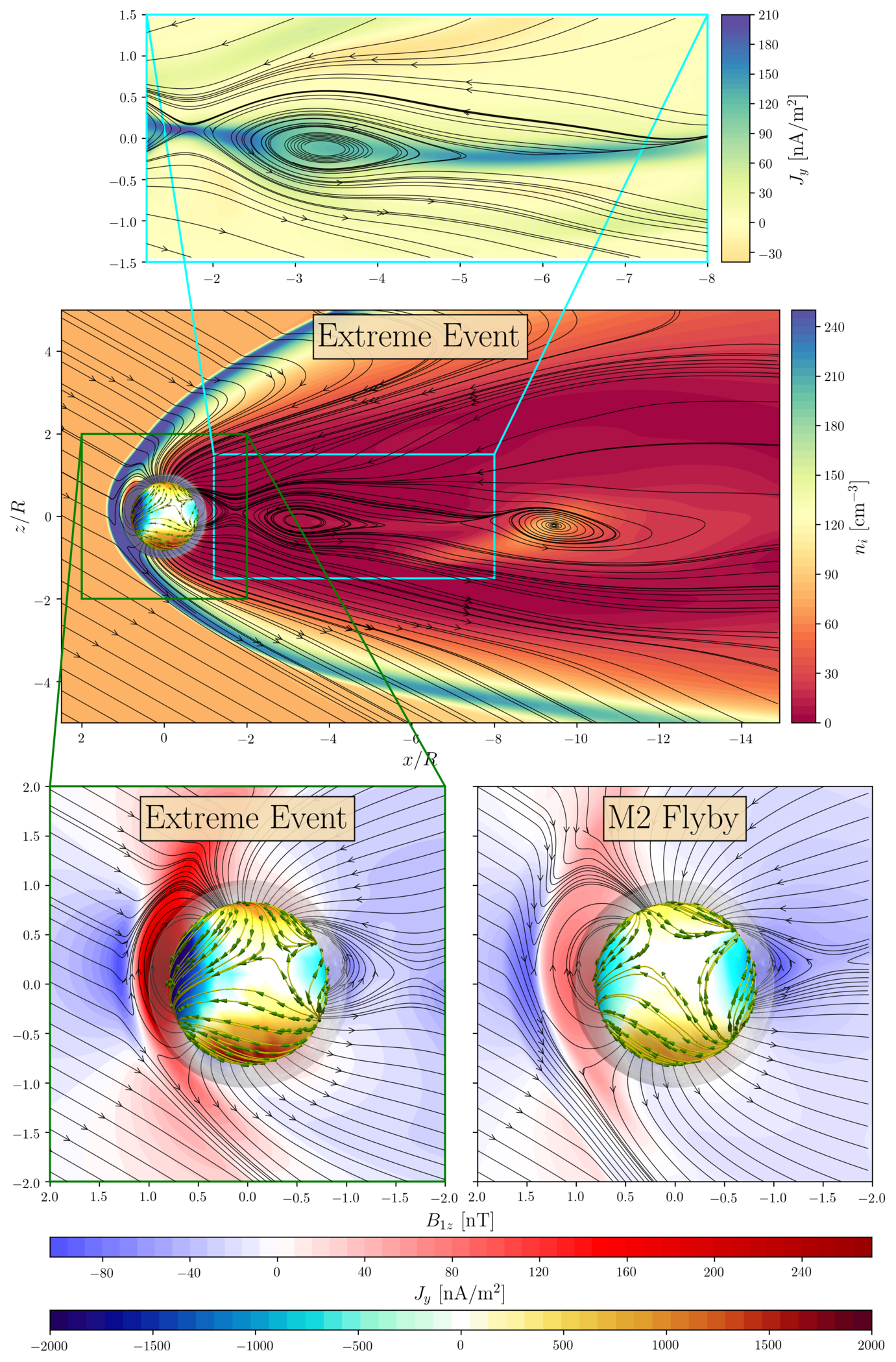}
\caption{Mercury's magnetosphere in the x-z (meridian) plane during a hypothetical extreme event. Plasmoids are formed in Mercury's magnetotail. The background color contours in the middle panel show the ion density in cm$^{-3}$. The bottom left panel shows the zoomed-in subdomain where color contours in the x-z plane represent the perturbation magnetic field $B_{1z}$ (in nT) and the color contours on the conducting core surface are the induction current $J_y$ (in nA/m$^2$). Note that the streamlines of core surface currents are illustrated by the yellow curves with green arrows wrapping around the core. Compared with the bottom right panel of M2, the $B_{1z}$ and the induction current $J_y$ from the extreme event are much stronger. The top panel depicts the formation of a plasmoid within the cross-tail current sheet.}
\label{extrem}
\end{figure}

The solar wind parameters of M2 yield a dynamic pressure of $\approx$11 nPa, which is relatively weak for instigating a significant induction response from the conducting core. Thus, we followed the scenario in \citet{jia15} to investigate the core-induced induction response; the solar wind density and speed are deliberately enhanced to 80 $cm^{-3}$ and 700 $km/s$, respectively, such that the solar wind dynamic pressure increases to $\approx$66 nPa, close to the pressure of 23 November 2011 event in \citet{slavin14}. The ten-moment multifluid calculation of Mercury's magnetospheric response to this hypothetical extreme event is shown in Figure \ref{extrem}. From the middle panel, one can see that both the bow shock and magnetopause boundaries are compressed significantly. Compared with the M2 flyby, the new magnetopause standoff distance is compressed to $\approx$1.15$R_M$, consistent with the results from \citet{jia15} for the same event study. In the bottom panel of Figure \ref{extrem}, we also compare the perturbation magnetic field $B_{1z}$ between the normal solar wind case (of M2) and the extreme event. As expected, solar wind compression increases $B_{1z}$ during the extreme event and squeeze the dayside magnetosphere. However, in order to demonstrate that the enhancement in $B_{1z}$ is not purely a result of solar wind compression, we present the core surface current $J_y$ for both cases, where the color contours on the core surface represent $J_y$ intensity and the yellow curves with green arrows are the corresponding current streamlines. Following Faraday's law of induction, these currents generate additional magnetic flux that acts against the solar wind pressure. By adopting the same color scale, it is clear that $J_y$ is much stronger in the extreme case than that in M2, indicating that the increase in $B_{1z}$ is a result of both solar wind compression and induction responses. The enhanced $B_{1z}$ and the intensified core surface current $J_y$ clearly demonstrate the importance of the induction response during the extreme event. 

In contrast to \citet{jia15}, our calculations contain richer features. For the first time, our simulation illustrates the formation of plasmoids in Mercury's magnetotail through \emph{collisionless} magnetic reconnection by including the reconnection electron physics. Plasmoids (or flux ropes) have, as a matter of fact, been observed by MESSENGER \citep{dibraccio15}. Theoretically speaking, these plasmoids are formed in elongated and intense current sheets due to the plasmoid instability - an explosive instability resulting in the formation of plasmoids due to magnetic reconnection \citep[e.g.,][]{comisso16}. In order to demonstrate that plasmoids are indeed formed within the cross-tail current layer, we plot the current sheet density ($J_y$) together with the plasmoid in the top panel of Figure \ref{extrem}. These plasmoids are eventually transported either toward or away from the planet, and new plasmoids will repeatedly form within the cross-tail current sheet (not shown here), leading to the small but extremely dynamic magnetosphere of Mercury. The impact of extreme space weather events (such as coronal mass ejections given in, e.g., \citealp{slavin14}) on Mercury's dynamic magnetosphere will be investigated in detail in our future work. 

\section{Conclusion}\label{sec:conclusion}

For the first time, we utilize a three-dimensional ten-moment multifluid model to study solar wind interaction with Mercury from the planetary interior to its dynamic magnetosphere. Given the importance of the induction effects shown in the previous studies, we also include a highly resistive mantle and an electrically conductive iron core (of radius 0.8$R_M$) inside the planet body. Direct comparison between MESSENGER magnetometer data and model calculations show good agreement, strongly supporting the validity of this new model. The cross-tail current sheet asymmetry revealed by the model is also consistent with MESSENGER observations. We conclude that the exhibited asymmetry in hot electron distribution is caused by the dual effect of Mercury's magnetotail reconnection and the dawnward drifts of electrons. In addition, this model accurately reproduces the field-aligned currents measured by MESSENGER that cannot be captured by an MHD model. Our study of magnetotail and magnetopause reconnection show that the off-diagonal elements of the electron pressure tensor, $\mathbf{P}_e$, play a key role in \emph{collisionless} magnetic reconnection. In order to investigate the induction effects, we have also studied Mercury's magnetospheric responses to a hypothetical extreme event. The simulation demonstrates that the induced magnetic fields help sustain a magnetopause, hindering the compression of the magnetopause down towards the surface. More interestingly, plasmoids (or flux ropes) are formed in Mercury's cross-tail current sheet, indicating Mercury's magnetotail being extremely dynamic. 

Thanks to this novel fluid approach that incorporates detailed electron physics associated with, e.g., \emph{collisionless} magnetic reconnection and magnetic drifts, we are able to reproduce and interpret the observations beyond MHD. Here we want to reiterate the distinction between the multi-moment multifluid approach and the (Hall) MHD approach from three perspectives. First, as mentioned earlier, the new model evolves the same set of equations (i.e., continuity, momentum and pressure tensor equations) for both ions and electrons (without the quasi-neutral assumption) and updates the electric and magnetic fields by adopting the full Maxwell's equations. As a result, the new model incorporates the non-ideal effects including the Hall effect, inertia, and tensorial pressures that are self-consistently embedded without the need for explicitly solving a generalized Ohm's law as MHD. Second, the new model supports all kinds of electromagnetic waves due to the inclusion of full Maxwell's equations. It is well-known that one of the shortcomings of Hall MHD lies in its failing to capture the right dispersion relation of Whistler waves (due to the assumption of massless electrons) when studying \emph{collisionless} magnetic reconnection. Last but not least, the new model contains an approximation to the Landau-fluid closure and therefore lower-order kinetic physics \citep{wang15,hammett90,hunana18}. For instance, the novel fluid approach can correctly capture the lower hybrid drift instability (LHDI), which can only be treated properly by a kinetic approach in the past \citep{ng19}.

In summary, MESSENGER furnished us with a great opportunity to study Mercury's dynamic magnetosphere. An abundance of useful data was returned from this mission, which stimulated numerous interesting studies. With the launch of the BepiColombo mission to Mercury in October 2018 \citep{benkhoff10}, Mercury's exploration will witness another notable surge after MESSENGER. A properly validated model that incorporates the electron physics essential for Mercury's \emph{collisionless} magnetosphere will likely advance our understanding of the dynamic responses of Mercury's magnetosphere to global solar wind interactions. Hence, the three-dimensional global ten-moment multifluid model developed herein represents a crucial step towards establishing a revolutionary approach that enables the investigation of Mercury's tightly coupled interior-magnetosphere system beyond the traditional fluid model, and has the potential to enhance the science returns of both the MESSENGER mission and the BepiColombo mission.

\acknowledgments
The authors thank Manasvi Lingam, Ryan Dewey, Suzanne Imber, Yuxi Chen, Yao Zhou, Chang Liu and Y. Y. Lau for the helpful discussions and comments. This work was supported by NSF Grant Nos. AGS-0962698 and AGS-1338944, NASA Grants Nos. 80NSSC19K0621, NNH13AW51I and 80NSSC18K0288 and DOE grant DE-SC0006670. The MESSENGER data used in this study are available from the PPI node of the Planetary Data System (\texttt{http://ppi.pds.nasa.gov}), and the model data were obtained from simulations using the \textsc{Gkeyll} framework developed at Princeton University, which is publicly available at \texttt{https://bitbucket.org/ammarhakim/gkeyll}. Resources supporting this work were provided by the NASA High-End Computing (HEC) Program through the NASA Advanced Supercomputing (NAS) Division at Ames Research Center, the Titan supercomputer at the Oak Ridge Leadership Computing Facility at the Oak Ridge National Laboratory through the INCITE program, supported by the Office of Science of the U.S. Department of Energy under Contract No. DE-AC05-00OR22725, the National Energy Research Scientific Computing Center, a DOE Office of Science User Facility supported by the Office of Science of the U.S. Department of Energy under Contract No. DE-AC02-05CH11231, Cheyenne (doi:10.5065/D6RX99HX) provided by NCAR's CISL, sponsored by NSF, and Trillian, a Cray XE6m-200 supercomputer at the UNH supported by the NSF MRI program under Grant No. PHY-1229408.






\bibliographystyle{agufull08}
\listofchanges

\end{document}